\documentclass[aps,twocolumn,prl,amsmath,amssymb,floatfix,superscriptaddress]{revtex4}

\usepackage{graphicx}
\usepackage{dcolumn}
\begin{document}

\bibliographystyle{revtex}


\title{Theory of the helical spin crystal: a candidate for the partially ordered state of MnSi}


\author{B. Binz}
\email{binzb@berkeley.edu} \affiliation{Department of Physics, University of California,  Berkeley, CA 94720, USA}
\author{A. Vishwanath}\affiliation{Department of Physics, University of California,  Berkeley, CA 94720, USA}
\author{V. Aji}
\affiliation{Department of Physics, University of California, Riverside, CA 92521, USA}


\date{\today}

\newcommand{\eref}[1]{(\ref{#1})}

\newcommand{\bv}[1]{{\bf #1}}
\newcommand{\uv}[1]{{\bf \hat #1}}

\newcommand{\be}{\begin{equation}}
\newcommand{\ee}{\end{equation}}

\begin{abstract}
MnSi is an itinerant magnet which at low temperatures develops a
helical spin density wave. Under pressure it undergoes a transition
into an unusual partially ordered state whose nature is debated.
Here we propose that the helical spin crystal (the magnetic analog
of a solid) is a useful starting point to understand partial order
in MnSi. We consider different helical spin crystals and determine
conditions under which they may be energetically favored. 
 The most promising candidate 
 has bcc structure 
 and is
reminiscent of the blue phase of liquid-crystals in that it has
line-nodes of magnetization protected by symmetry. 
 We introduce a
Landau theory to study the properties of these states, in particular
the effect of crystal anisotropy, magnetic field and disorder. These
results compare favorably with existing data on MnSi from neutron
scattering and magnetic field studies. Future experiments to test
this scenario are also proposed.

\end{abstract}

\maketitle

MnSi is a well studied itinerant ferromagnet \cite{moriya} but has
recently been at the focus of renewed attention since the discovery
of unusual properties in high pressure studies
\cite{thessieu97,pfleiderer01,pfleiderer04,NMR}.
The ambient-pressure magnetic phase of MnSi is characterized by a
hierarchy of three major energy scales \cite{nakanishi80}.
First are interactions that favor itinerant ferromagnetism. At a
much lower energy scale, Dzyaloshinskii-Moriya (D-M) spin-orbit
coupling (which is allowed in the non-centrosymmetric B20 crystal
structure of MnSi) leads to a spiraling of the magnetic moment  with a single helicity (helical
spin-density wave). The
small ratio between these scales is responsible for the long spiral pitch
 ($\lambda=180\mathring{A}$ vs. the lattice
constant $a=4.6\mathring{A}$).
The D-M interaction defines chirality and the length of the ordering
wave-vector ($Q=2\pi/\lambda$)  but not its direction. The latter is
pinned to $\langle111\rangle$ by even weaker crystal-anisotropy terms [$\langle111\rangle$: class of crystal directions related by cubic symmetry].
Elastic neutron scattering at atmospheric pressure reveals Bragg
spots at these eight points in the Brillouin zone which are
attributed to domains each containing a single spiral state. On
application of pressure however, a first order phase transition
occurs at $p_c=14.6{\rm kbar}$. At higher pressures, unusual
properties are observed
. First, the neutron scattering signal is completely
changed from the low pressure phase. Enhanced scattering is seen
at wavevectors with a length similar to the low pressure phase, but
with intensities no longer sharply peaked along $\langle111\rangle$. 
Rather, the intensity is more 
diffuse over the
wavevector sphere, hence the name 
"partial order", 
 but clearly peaking at $\langle110\rangle$
\cite{pfleiderer04}. Secondly, the resistivity displays a non fermi
liquid (NFL) temperature dependence 
\cite{pfleiderer01}.
Although they both onset at $p_c$, the NFL persists to pressures far
beyond where "partial order" is seen. The relation between these two
puzzles is therefore unclear - here we focus only on offering a
theory for the "partially ordered" state. Recent theoretical work on
collective modes and electronic properties of helimagnets 
are in Refs. \cite{theory}.
Other theoretical
proposals for the high pressure state of MnSi have invoked proximity to a
quantum multi-critical point \cite{schmalian04}, magnetic liquid-gas transitions
\cite{tewari05} 
 and skyrmion-like structures
\cite{bogdanov05,fischer06}.

None of these theories has attempted to explain the
new peak positions. As a first attempt, one might
speculate that the crystal anisotropy which pins the spiral along 
$\langle111\rangle$
 at low pressure is modified at high pressure and
pins it along a different set of directions, $\langle110\rangle$. However,
this view is untenable on many counts, e.g. it is at odds with the
magnetic field studies \cite{thessieu97}. Moreover, the usual
crystal anisotropy terms consistent with the lattice symmetry give
energy minima either at $\langle111\rangle$ or $\langle100\rangle$; but $\langle110\rangle$ are
always saddle points (which can be proved more generally
\cite{long_paper}). Similarly, while it is tempting to interpret the
partially ordered phase of MnSi entirely in terms of a directionally
fluctuating spin spirals, this requires both a novel theoretical
framework for its description, as well as an explanation for the
unusual anisotropy.

In this work, we study ordered magnetic states which are linear
superpositions of spin spirals with different wave-vectors.
We show that in addition to the three energy scales (ferromagnetic,
D-M and crystal anisotropy) that are usually considered in MnSi,
there is an additional scale that is important, the interaction
between different modes (expected to be intermediate between the D-M
and crystal anisotropy), which determines the spin crystal state
that is selected. We propose that the transition under pressure
between the single-spiral and partially ordered states is driven by a
change in the inter-mode interactions that goes from preferring a
single-spiral ground state to a multi-spiral ground state.

{\em Magnetic weak crystallization theory:} The Landau free energy
 to second order in the local spin density $\bv M(\bv r)$ of a 
system with full rotation symmetry (transforming both
space and spin together) - but no inversion symmetry is  \be F_2=\left\langle
r_0 \bv M^2+J(\nabla \bv M)^2+2D\,\bv M\cdot(\nabla\times \bv
M)\right\rangle, \label{F2} \ee where $\langle\ldots\rangle$ means
averaging over the sample and $r_0,J,D$ are parameters ($J>0$).
Coupling to fermions which leads to damped dynamics is ignored in
the following since we will focus on ground states with long range
order. The last term of Eq.~\eref{F2} is the D-M interaction.
Clearly, the energy is minimal for circularly polarized  waves of
fixed helicity which satisfy $D\,\bv m^*_{\bv q}\cdot (i\bv q\times
\bv m_{\bv q})=-|D|\,q\,|\bv
m_{\bv q}|^2$, where $\bv m_{\bv q}$ are the Fourier modes of $\bv M(\bv r)$. 
Equation \eref{F2}, then becomes $F_2=\sum_{\bv q}r(q)\,|\bv m_{\bv
q}|^2$, where $r(q)=r_0-D^2/J+J(q-|D|/J)^2$. Here, in analogy to
standard liquids (and in contrast to the uniform ferromagnetic
case),
$r(q)$ is minimized at a finite wavevector $Q=|D|/J$ and for
$r(Q)\to0$, all modes on the surface $|\bv q|=Q$ in reciprocal space
become soft (as opposed to a single point $\bv q=0$). To study the
implications of such a singular surface on the phase transition
itself is beyond the scope of this letter. But we recognize that
when $r(Q)\to0$, the Gaussian theory not only becomes unstable
towards formation of a helical spin-density wave along any
direction, but that linear combinations of such states (helical spin
crystals) are equally natural candidates.

In the spirit of weak crystallization theory \cite{brazovskii87}, we
now study  minima of the free energy in the ordered phase, where
$r(Q)<0$. The degeneracy between a simple spin spiral and linear
combinations of several spin spirals (spin crystals) is lifted by
interactions  i.e. by the fourth-order term ($F_4$) in $\bv M$ (odd
terms in $\bv M$ are forbidden by time reversal symmetry). We assume
that $F_4$, as $F_2$, has full rotation symmetry and we will include
the weak crystal anisotropy last. Hence, with $\bv q_4=-(\bv q_1+\bv
q_2+\bv q_3)$: \be F_4=\!\!\sum_{\bv q_1,\bv q_2,\bv q_3}\!\!U(\bv
q_1,\bv q_2,\bv q_3)\left(\bv m_{\bv q_1}\cdot\bv m_{\bv
q_2}\right)\left(\bv m_{\bv q_3}\cdot\bv m_{\bv
q_4}\right).\label{F4} \ee  If the interaction is strictly local in
real space [$F_4\propto\langle\bv M^4\rangle$, which is equivalent to a
constant coupling $U(\bv q_1,\bv q_2,\bv q_3)=U_0$], it is easy to
show that the single-mode spin spiral is the absolute minimum of
$F_2+F_4$. In the spin spiral state, $\bv M^2$ is constant and it
is therefore possible to minimize both $F_2$ and $F_4$
independently.

However, $U(\bv q_1,\bv q_2,\bv q_3)$ is generally not constant.
Assuming $F_4$ is small, in that its main effect is to provide an
interaction between the modes which are degenerate under $F_2$, the
important terms of $F_4$ are those with $|\bv q_1|= |\bv q_2|=|\bv
q_3|=|\bv q_4|=Q$. Then the
coupling function $U$ depends only on two independent angles
$2\theta=\arccos(\uv q_1\cdot\uv q_2)$ and
$\phi/2=\arccos[(\uv q_2-\uv q_1)\cdot \uv q_3\,/(1-\uv
q_1\cdot\uv q_2)]$
\footnote{Geometrically, 
$\phi/2$ is the angle between the two planes
spanned by $(\bv q_1, \bv q_2)$ and  $(\bv q_3,\bv q_4)$. If $\bv q_1+\bv q_2=0$, $\phi/2$
 is the angle between $\bv q_2$ and  $\bv q_3$.}.
 This mapping allows $\theta$ and $\phi$ to be interpreted as the
polar and azimuthal angles of a sphere and the coupling
$U(\theta,\phi)$ is a smooth function on that sphere satisfying:
$U(\theta,\phi)=U(\pi-\theta,\phi)=U(\theta,2\pi-\phi)$. It is
natural to expand $U$ in spherical harmonics ($Y_{lm}$)
which satisfy this relation and 
we just retain terms with
$l\leq2$. Thus
$U(\theta,\phi)=U_0+U_{11}\sin\theta\cos\phi+U_{20}(3\cos^2\theta-1)+U_{22}\sin^2\theta\cos2\phi$.

\begin{figure}
\includegraphics[scale=0.6]{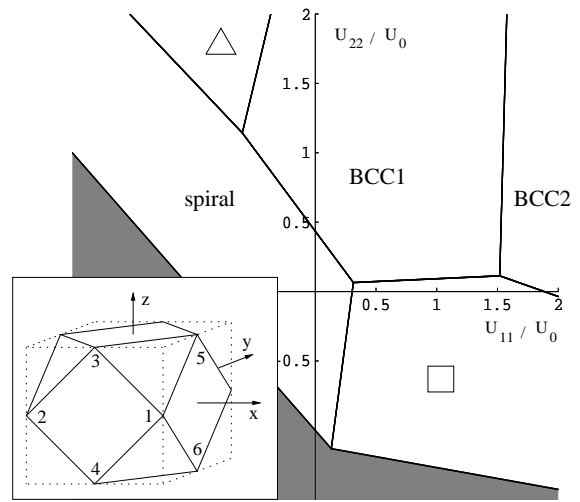}
\caption{Mean-field phase diagram of the six-mode model for $r(Q)<0$
as a function of the interaction parameters for $U_{20}=0$ and
$U_0>0$. In the gray region, $F_4<0$ and higher-order terms should
be added for stability. \label{ylm-diag}}
\end{figure}

To find the absolute minimum of $F_2+F_4$ for a general
angle-dependent interaction would be a  formidable task.
Instead, we first
limit ourselves to linear combinations of six spin
spirals, corresponding to wavevectors $\pm\bv k_j$ (for $j=1,\ldots,6$),
where $\bv k_1=Q/\sqrt2\,(1,-1,0)$,  $\bv k_2=Q/\sqrt2\,(-1,-1,0)$, etc. (see inset of Fig.~\ref{ylm-diag}).
We may write
the six modes as $\bv m_{\bv k_j}=\bv m^*_{-\bv k_j}=\psi_j\, \bv n_{\uv k_j}$, where $\bv n_{\uv q}=[\uv q\times(\uv  z\times\uv
q)+i\uv z\times\uv q]/[1-(\uv z\cdot\uv q)^2]^{1/2}$. The choice of
the unit vector $\uv z$ is arbitrary. A different choice leads to a set of phase changes in the six complex variables
$\psi_j$. The interaction Eq.~\ref{F4}, written in terms of the $\psi$-variables, contains
terms of the form $V_{jj'}|\psi_j|^2|\psi_{j'}|^2$ (three parameters)
and one term  $\lambda\, \Re \left(T_x+T_y+T_z
\right)$, where
$T_x=\psi_1^*\psi_2\psi_5\psi_6$,
$T_y=\psi_1^*\psi_2^*\psi_3\psi_4$ and
$T_z=-\psi_3\psi_4^*\psi_5^*\psi_6$ which is sensitive to relative phases. These are the only quartic terms invariant under the microscopic symmetries of the problem: translations, time reversal and tetrahedral point group operations. The parameters $V_{jj'}$ and $\lambda$  can be easily expressed as a linear
combination of $U_0, U_{11}, U_{20}$ and $U_{22}$.
The resulting mean-field phase
diagram is shown in
Fig.~\ref{ylm-diag}. Different ordered phases are separated by
first-order phase boundaries. In the  
"spiral" state,
only one out of six amplitude  is non-zero. The square lattice state
``$\square$'' is an superposition of two orthogonal spin spirals with equal
amplitude.
In the region ``$\bigtriangleup$'', there are actually two
degenerate states, one with $|\psi_i|>0$ ($i=1,\,3,\,5$)
and one with ($i=2,\,4,\,6$)
(remaining amplitudes being zero in
each case).  Finally in BCC1 ($\lambda>0$) and BCC2 ($\lambda<0$), all six modes contribute with
equal amplitudes. These states have the 
 periodicity of a BCC crystal and will be the focus of this paper.
Note that a small correction of the purely local interaction $U_0$,
given by $U_{11}\approx U_{22}\approx 0.2U_0$ is sufficient to
induce a transition from the spiral state to the BCC1 spin crystal.
\footnote{We have also checked the
energies of some spin crystal states outside of the six-mode model
 \cite{long_paper}. The transition from the spin spiral to the BCC1 state remains stable
against all rivaling states tested so far.}

{\em Properties of the BCC spin crystal:}
From the sole assumption (suggested by experiments) that a spin texture has
equal-weighted Bragg peaks at the set of wave-vectors $\pm\bv k_j$,
weak crystallization theory restricts an infinite variety of
physically very different magnetization patterns $\bv M(\bv r)$ down
to only two possibilities: BCC1 and BCC2. In BCC1 (BCC2),  the relative phases 
of $\psi_1,\ldots,\psi_6$ are locked in such a way that $T_x,T_y$ and $T_z$ are all 
negative (positive), 
i.e. for BCC1: $\hat\psi_1=\pm\hat\psi_4\hat\psi_5$, $\hat \psi_2=\mp\hat\psi_4\hat\psi_6^*$, $\hat\psi_3=\hat\psi_4\hat\psi_5\hat\psi_6^*$ where $\hat\psi_j=\psi_j/|\psi_j|$ and for BCC2: $\hat\psi_1=\pm i\hat\psi_4\hat\psi_5$, $\hat \psi_2=\pm i\hat\psi_4\hat\psi_6^*$, $\hat\psi_3=-\hat\psi_4\hat\psi_5\hat\psi_6^*$.  
Three phases (e.g. $\hat\psi_4,\hat\psi_5$, $\hat\psi_6$)
are arbitrary due to global translation symmetry. In addition to the translational
degeneracy, both BCC states are two-fold degenerate due to
time-reversal symmetry breaking.  In contrast to the single spiral,
time-reversal [$\bv M(\bv r)\to-\bv M(\bv r)$] is {\it not}
equivalent to a translation in either of the BCC states. In fact,
both BCC states feature a macroscopic time-reversal symmetry
breaking order parameter $S=\langle M_xM_yM_z\rangle\neq0$.

\begin{figure}
\includegraphics[scale=0.5]{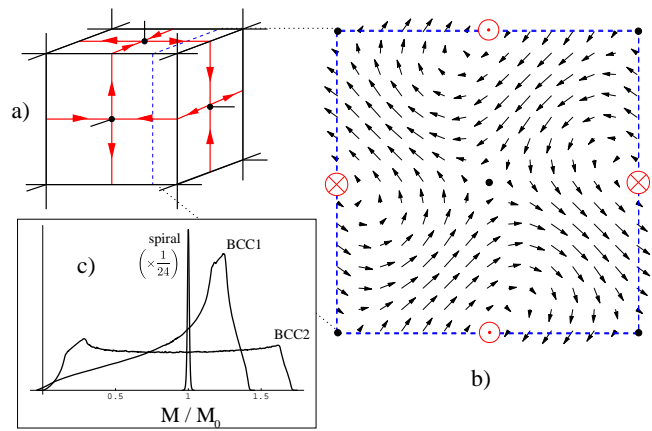}
\caption{(Color online) (a) Magnetization pattern $\bv M(\bv r)$ of BCC1. The black lines are nodes, where $\bv M=0$. Along red lines, the magnetization direction is as indicated. The blue dashed line indicates the location of the cut shown in (b) where vectors denote the in-plane magnetization. The nodes lines are the centers of anti-vortices and the directed red lines are the centers of meron configurations.
(c) Probability distributions $\langle\delta(M-|\bv M(\bv
r)|)\rangle$ of three different magnetic states: a single-spiral
state of amplitude $M_0$ and both BCC states with
$|\psi_j|^2=M_0^2/6$. Landau theory predicts this $1/6$ ratio of
amplitudes at the phase boundary between these states.
A finite-width $\delta$-function was used for the numerical evaluation.
\label{real}}
\end{figure}

BCC1 can be {\em defined} by its symmetry properties as being the unique structure that is invariant under time reversal (sign change of $\bv M$) followed by a $\pi/2$ rotation about the  $x$, $y$ or $z$ axis.
For BCC2, the same operations result in translations.
 The real space BCC1 structure is illustrated in
Fig.~\ref{real}. There are two sets of straight lines, along which
$\bv M$ is constrained by the symmetry properties of BCC1. First, the
magnetization vanishes along the $x$, $y$ and $z$ axes [and their
translations by 
$(\frac12,\frac12,\frac12)$], due to the invariance
mentioned above. These are reminescent of the blue phases of chiral
nematic liquid crystals \cite{mermin}, where line defects also exist but
are protected by topology, rather than by symmetry. Second, in
the center between four parallel node lines, the magnetization direction is
constrained as shown in
Fig.~\ref{real} (a). Since these properties are determined by
symmetry, they are stable, even if higher Fourier modes are
included. In contrast, BCC2 has no node lines, but six point nodes in
each primitive unit cell.  Fig.~\ref{real} (c) shows
the distribution of the local magnetization $|\bv M|$ over the sample.

{\em Anisotropy and locking of crystal directions:}
So far, our model free energy [Eqs. \eref{F2} and \eref{F4}] has
been completely rotation invariant. By choosing those six modes
which lead to a BCC crystal, we assume that full rotation symmetry
is spontaneously broken, but a global rotation 
still leaves the energy invariant.
This degeneracy is broken by small anisotropic terms.
 For the MnSi B20 crystal structure, the leading anisotropic term in powers of $M$ and
$q$ is of the form: $ F_a=a \sum_{j}g({\uv k_j})|\psi_j|^2$ where
$g(\uv k)=\hat k_x^4+\hat k_y^4+\hat k_z^4$. The function $g$ has
its maxima and minima along $\langle100\rangle$ and $\langle111\rangle$,
respectively, while $\langle110\rangle$ is a saddle point. It is therefore
impossible that this term locks a single-spiral to $\langle110\rangle$. 
From the observed spiral orientation along $\langle111\rangle$ at low
pressures, we obtain $a>0$. Assuming there is no sign change with
pressure, we find that this locks 
the magnetic BCC crystal (BCC1 or BCC2) to the atomic
B20 crystal in such a way that all $\uv k_j$ point along $\langle110\rangle$. 
Hence it naturally explains the neutron scattering peaks
at the locations seen in Ref.~\cite{pfleiderer04}.

{\em Magnetic field:}
A uniform magnetic field couples linearly to the $\bv q=0$ mode of the
magnetization, $\bv m=\langle\bv M\rangle$. Eq.~\eref{F4} couples
this additional mode  to the spin crystal modes and produces terms of the form $\bv m^2\,|\psi_j|^2$,
$(\bv m\cdot\uv k_j)^2|\psi_j|^2$ and one term $\mu \,\bv h_\psi\cdot\bv m$, where $\bv h_\psi$ is cubic in the $\psi$ variables \cite{long_paper}.
 The behavior of 
 a single spin spiral
 in a magnetic field is characterized by a strongly anisotropic susceptibility,
 which effectively orients
the spiral axis along the field  \cite{kataoka81}. Once the spiral is oriented, the
susceptibility is comparatively large. In the BCC spin crystal
states, there is no such anisotropy in the linear response, because
 symmetry  does not allow for an anisotropic
susceptibility tensor.
Since the BCC spirals, unlike the single spiral,  cannot be optimally oriented along the field,
 one expects the following sequence of
phases as a function of increasing magnetic fields: ``BCC'' $\to$
``spiral'' $\to$ ``spin polarized'', in good comparison with
experiment \cite{thessieu97}. This expectation is true within our
theory if 
$\mu$ is not too large. In fact, the term
$\mu \,\bv h_\psi\cdot\bv m$ induces a change of the relative
amplitudes and phases of the six interfering spirals as a function
of the magnetic field and therefore adds to the ability of the BCC
state to adjust to an external field. For example, a magnetic field
in $z$-direction suppresses $|\psi_1|$ and enhances $|\psi_2|$
relative to the other four amplitudes [or the opposite, depending on
the signs of $\mu$ and the time-reversal index $S=\langle M_xM_yM_z\rangle$].
  This effect should be observable by neutron scattering in a single-domain sample,
    which may  be
obtained via proper field-cooling. In practice this needs some care
since the energy splitting between the two time reversed states is
only cubic in the field: $\propto H_x H_y H_z$.

{\em Magneto-transport:} The broken time-reversal symmetry in the
BCC spin crystals, but with the absence of a uniform magnetization,
can give rise to unusual magneto-transport in single-domain samples.
The symmetric part of the conductivity tensor is allowed a linear
field dependence due to time reversal symmetry breaking ($S\neq0$):
\be \sigma_{ab}=\sigma_0\delta_{ab}+\alpha \, S\, |\epsilon_{abc}|
H_c+O\left(H^2\right). \ee
E.g. for a field applied along $\uv z$, the conductivity
along  $(\uv x\pm\uv y)/\sqrt{2}$ would display {\em linear}
magneto-resistance. Similarly, the Hall conductivity (the
antisymmetric part of $\sigma$) is allowed an unusual quadratic
contribution: \be \sigma_{ab}^H=\sigma^H_0\epsilon_{abc}H_c+\alpha'
S\, \epsilon_{abc}|\epsilon_{cde}|H_dH_e+O\left(H^3\right). \ee
Thus, the Hall effect will in general not simply switch sign if the
direction of the magnetic field is inverted.

{\em Effect of Disorder:} Although the
available MnSi
crystals are very clean from the electrical resistivity point of
view, 
helical magnetic structures are sensitive to disorder at a
much longer length scale.
 Hence disorder effects need to be studied. Non-magnetic disorder which couples to the
magnetization squared, $F_{dis}=\int d{\bv r}\; V_{dis}(\bv r) |\bv
M(\bv r)|^2$, will not affect single-spiral states which have a constant magnitude of magnetization, but they do affect spin crystal states which have a modulated
magnitude  [see Fig \ref{real} (c)].
Therefore the neutron
scattering signal of the spin-crystal state is expected to have more
diffuse scattering than the single spiral,
consistent with the
experimental observation that the high pressure phase has diffuse scattering peaked about $\langle110\rangle$ while the low pressure phase
has sharper spots. Anisotropic spreading of the Bragg spots parallel
and perpendicular to the wavevector sphere is also anticipated
\cite{long_paper}. Disorder in D=3 is expected to destroy true long
range order of the spin crystal, leading to either a Bragg glass or a
disordered state. In either case, time reversal symmetry breaking of
BCC1 remains, implying that there must be a finite temperature phase
transition on cooling into this phase.

{\em NMR and $\mu$SR:}
The spatially modulated magnetization magnitude of spin-crystal states  [Fig.~\ref{real} (c)]
should be visible in muon spin rotation ($\mu$SR) 
and 
zero field NMR experiments.
While $\mu$SR 
on MnSi 
has 
been performed only at low pressures \cite{muSR},
 zero field NMR 
 was carried out over a wide range of pressure \cite{NMR}.
 The resonant
frequency was found to be sharp in the low pressure phase but to
broaden 
at higher pressure
.
Both the
presence of static magnetism and the broader distribution of
frequencies ($|\bv M|$ distribution) is consistent with our proposal of
BCC1 for the high-pressure phase.

{\em Conclusions:} A 
time reversal symmetry breaking %
helical spin crystal BCC1, disordered by weak
impurities
, is proposed for the high pressure
'partially ordered' state of MnSi. Both theoretical arguments from
simple models where this state naturally arises adjacent to the
single spiral state, as well as favorable comparison with a variety
of experiments: e.g. neutron scattering and magnetic field studies
lends support to this view. Future experiments that could probe the
unusual properties of this phase e.g. magneto-conductivity and
$\mu$SR are also proposed. Important puzzles that remain are the
origin of 
NFL behavior 
and the 
 lack of clear signatures for a 
finite temperature transition into the ``partially
ordered'' state.

\acknowledgments
We thank L. Balents, I. Fischer, D. Huse, J. Moore, M.P. Ong, C. Pfleiderer,   D. Podolsky, A. Rosch and T. Senthil for
stimulating discussions; the Swiss National Science Foundation
(B. B.), the A.P. Sloan Foundation and DE-AC02-05CH11231 (A.V.) for
financial support.

\end{document}